\documentclass[journal, 12pt]{IEEEtran}
\onecolumn
\usepackage{setspace}
\usepackage{amsmath}
\usepackage{amsthm}   
\usepackage{amssymb} 
\usepackage[bottom,hang,flushmargin]{footmisc}
\usepackage{graphicx}
\usepackage{caption}
\usepackage{algorithm}
\usepackage{algorithmic}
\usepackage{hyperref}

\renewcommand{\footnoterule}{%
  \kern -3pt
  \hrule width 0.4\textwidth height 0.4pt
  \kern 2.6pt
}

\begin{document}

\title{An Innovative Networks in Federated Learning}

\author{Zavareh Bozorgasl,~\IEEEmembership{Member,~IEEE}, Hao~Chen,~\IEEEmembership{Member,~IEEE}  \thanks{Z. Bozorgasl (zavarehbozorgasl@boisestate.edu) and H. Chen (haochen@boisestate.edu) are with the Department of Electrical and Computer Engineering, Boise State University, Boise,
ID, 83712.} }

\maketitle

\begin{abstract}
This paper presents the development and application of Wavelet Kolmogorov-Arnold Networks (Wav-KAN) in federated learning. We implemented Wav-KAN  \cite{wav-kan} in the clients. Indeed, we have considered both continuous wavelet transform (CWT) and also discrete wavelet transform (DWT) to enable multiresolution capabaility which helps in heteregeneous data distribution across clients. Extensive experiments were conducted on different datasets, demonstrating Wav-KAN's superior performance in terms of interpretability, computational speed, training and test accuracy. Our federated learning algorithm integrates wavelet-based activation functions, parameterized by weight, scale, and translation, to enhance local and global model performance. Results show significant improvements in computational efficiency, robustness, and accuracy, highlighting the effectiveness of wavelet selection in scalable neural network design.
\end{abstract}


\IEEEpeerreviewmaketitle

\section{Introduction}

\textbf{Remark}: The results in the simulation section consist of i.i.d. data distribution for the MNIST dataset. We have considered different types of wavelets in Wav-KAN \cite{wav-kan} in federated learning to solve i.i.d., unbalanced, and non-i.i.d. data distribution of different datasets (including CIFAR10, CIFAR100, FEMNIST, and CelebA). We will provide the analysis for the case of both CWT KAN, CWT KAN with integer scale and translation, and also DWT KAN (which is excellent for multiresolution analysis) on each client in the federated learning setting. Moreover, more analyses and proofs, including the convergence for both homogeneous and heterogeneous data, have been proven and we will publish them in the next versions\footnote{Codes will be available at \href{https://github.com/zavareh1}{https://github.com/zavareh1}}. \\

\section{FL-Wav-KAN: Federated Learning based on Wavelet Kolmogorov-Arnold Networks}
In this federated learning algorithm, we incorporate the use of different mother wavelet \footnote{We publish an extensive result that we have taken.} mother wavelet for the learnable activation functions in each client node. Each activation function at the client side is parameterized by three parameters: weight, scale, and translation. These parameters are crucial for the wavelet-based activation functions to operate effectively. In the following we discuss Mexican hat wavelet as an example.
Wavelets equivalent to the second derivative of a Gaussian, known as \textit{Mexican hats}, were initially utilized in computer vision for detecting multiscale edges \cite{witkin,mallat}. They have the following form

\begin{equation}
    \psi(t) = \frac{2}{\pi^{1/4} \sqrt{3\sigma}} \left( \frac{t^2}{\sigma^2} - 1 \right) \exp \left( \frac{-t^2}{2\sigma^2} \right). 
\end{equation}

Where $\sigma$ shows the adjustable standard deviation of Gaussian. In our experiments, $\psi_{exp}(t)$ 
\begin{equation}
  \psi_{exp}(t) = w \psi(t)  
\end{equation}
 Indeed, $w$ plays the role of CWT coefficients which is multiplied by the mother wavelet formula; as $w$ is a learnable parameter, it helps adapting the shape of the mother wavelet to the function that it tries to approximate.

\begin{algorithm}[htbp]
\caption{Federated Learning with Wav-KAN}
\label{alg:fed_wavkan}
\begin{algorithmic}[1]
\STATE Each client downloads a neural network with the same structure and reports the size of their local dataset to the server.
\FOR{$t = 1$ to $T$}
    \FOR{each client selected by the server for this round}
        \FOR{$K$ training steps/epochs}
            \STATE Train the neural network on the local dataset using a mother wavelet like the Mexican hat wavelet for activation functions.
        \ENDFOR
        \STATE Send updated parameters (weights, scale, and translation for each activation function) to the server.
    \ENDFOR
    \STATE Server aggregates parameters by averaging:
\[
\bar{w} = \frac{1}{N} \sum_{i=1}^{N} w_i \quad
\bar{s} = \frac{1}{N} \sum_{i=1}^{N} s_i \quad
\bar{\tau} = \frac{1}{N} \sum_{i=1}^{N} \tau_i
\]
    \STATE Server broadcasts aggregated parameters to all clients.
    \STATE Clients update local models using aggregated parameters.
\ENDFOR
\end{algorithmic}
\end{algorithm}

As we see in Table \ref{alg:fed_wavkan}, Initially, every client downloads a neural network with an identical structure and reports the size of their local dataset to the server. During each training round, the selected clients perform the following steps:

1. Training Phase: For a predefined number of training steps or epochs, each client trains the neural network on their local dataset. For each neuron (edge), the activation function appears as:

\begin{equation}
    \psi(\frac{t-\tau_i}{s_i}) = \frac{2}{\pi^{1/4} \sqrt{3\sigma}} \left( \frac{(\frac{t-\tau_i}{s_i}))^2}{\sigma^2} - 1 \right) \exp \left( \frac{-(\frac{t-\tau_i}{s_i})^2}{2\sigma^2} \right). 
\end{equation}

Indeed, at first, we initialize these parameters and clients download the initial model with the initialized parameters, including wavelet coefficients, translations and scales. Then, each neuron affects its input, and the learnable parameters will be updated by backpropagation.  

2. Parameter Transmission: After completing the local training in a predefined number of iterations, each client sends the updated parameters (weights, scale, and translation for each activation function) to the server. 

3. Aggregation at the Server: The server aggregates the received parameters by averaging them. Specifically, the server computes the average weight, scale, and translation parameters across all clients. The average of these parameters is calculated as follows:

   \[
\bar{w} = \frac{1}{N} \sum_{i=1}^{N} w_i
\]
\[
\bar{s} = \frac{1}{N} \sum_{i=1}^{N} s_i
\]
\[
\bar{\tau} = \frac{1}{N} \sum_{i=1}^{N} \tau_i
\]

    Here, \( w_i \), \( s_i \), and \( t_i \) represent the weight, scale, and translation parameters from each client, and \( N \) is the number of clients.

4. Broadcasting the Global Model: The server then sends the aggregated parameters back to all clients. In the subsequent training round, each client updates their local model using these aggregated parameters, ensuring consistency and improved performance across the federated network.

This process is repeated for a predefined number of rounds, \( T \), allowing the global model to iteratively improve through the collaborative training efforts of all clients. We will discuss different cases of model updating in the future versions of this work.

\section{Simulation Results}
\label{sec:sim}

In this section, we present experiments using wav-KAN model with various wavelet transformations on the MNIST dataset. Our dataset consists of a training set of 50K images and a test set of 10K images. The primary aim of this experiments was not to achieve optimal parameter values but to illustrate the effectiveness of Wav-KAN in terms of overall performance. A lot of mother wavelet types examined  including Mexican hat, Morlet, Derivative of Gaussian (DOG), and Shannon wavelets, as summarized in Table \ref{tab:mother_wavelets}. Also, we have implemented the federated learning version of Spl-KAN which both in terms of communication and computation is weaker than Wav-KAN (refer to Table I of Wav-KAN paper).

For each wavelet type, we conducted 3 trials, training the model for 50 epochs in each trial. To guarantee the robustness and reliability of our findings, the results were averaged over multiple trials. The Wav-KAN models were configured with nodes in the arrangement [28*28, 64, 10]. We used the AdamW optimizer with a learning rate of 0.001 and a weight decay of $10^{-4}$, and cross-entropy was employed as the loss function.

\begin{table}[htbp]
    \centering
    \caption{Mother Wavelet Formulas and Parameters}
    \label{tab:mother_wavelets}
    \begin{tabular}{|c|c|c|}
        \hline
        \textbf{Wavelet Type} & \textbf{Formula of Mother Wavelet} & \textbf{Parameters} \\
        \hline
        Mexican hat & $\psi(t) = \frac{2}{\sqrt{3}\pi^{1/4}} \left(t^2 -1\right) e^{-\frac{t^2}{2}}$ &  $\tau$, $s$  \\
        \hline
       Morlet & $\psi(t) = \cos(\omega_0 t) e^{-\frac{t^2}{2}}$ & $\tau$, $s$ and scaling, $\omega_0$= 5 \\
        \hline
        Derivative of Gaussian (DOG) & $\psi(t) = -\frac{d}{dt}\left( e^{-\frac{t^2}{2}} \right)$ & $\tau$, $s$ \\
        \hline
        Shannon & $\psi(t) = \text{sinc}(t/\pi) \cdot w(t)$ & $\tau$, $s$, and $w(t)$: window function \\
        \hline
    \end{tabular}
\end{table}

\begin{figure}[ht]
    \centering
    \includegraphics[width=0.8\linewidth]{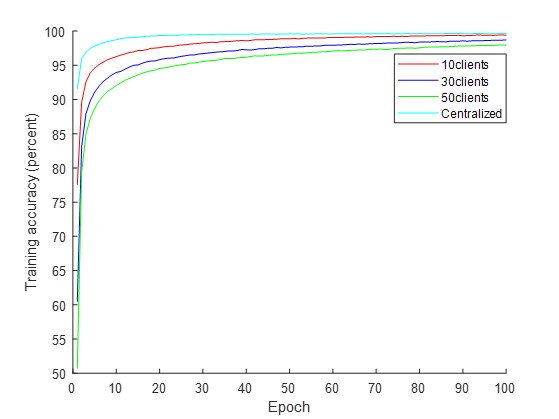} 
    \captionsetup{justification=centering} 
    \caption{Training accuracy of Wav-KAN [28*28,64,10] with Mexican hat mother wavelet activation.} 
    \label{fig:MNIST_train_acc}
\end{figure}

\begin{figure}[ht]
    \centering
    \includegraphics[width=0.8\linewidth]{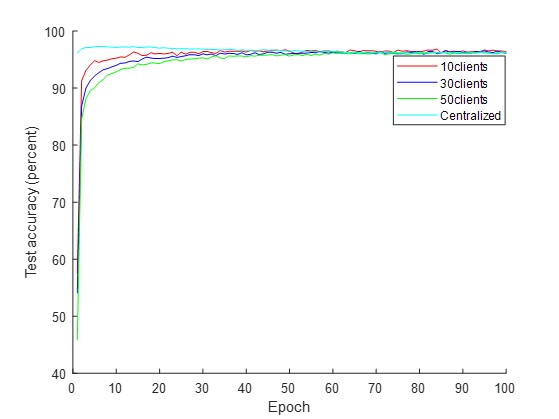} 
    \captionsetup{justification=centering} 
    \caption{Test accuracy of Wav-KAN [28*28,64,10] with Mexican hat mother wavelet activation.} 
    \label{fig:MNIST_test_acc}
\end{figure}

Figures \ref{fig:MNIST_train_acc} and \ref{fig:MNIST_test_acc} display the training and test accuracy of Wav-KAN with Mexican hat wavelets averaged for three trials for each wavelet type. By fine-tuning and utilizing the flexibility of wavelets (such as the frequency of the sinusoid and the variance of the Gaussian), wavelets showed significant superiority. In this experiment, we set the Gaussian variance to 1; however, better variances could be found through grid search or by making it a learnable parameter. Wavelets effectively balance by avoiding overfitting to noise in the data.

The simulation results highlight the significant impact of wavelet selection on the performance of the KAN model. This suggests that certain wavelets are particularly effective at capturing the essential features of the dataset while maintaining robustness against noise. Conversely, wavelets like Shannon and Bump did not perform as well, emphasizing the importance of wavelet selection in designing neural networks with wavelet transformations.

\section{Conclusion}
\label{sec:conclusion}

In this paper, we introduced the Wavelet Kolmogorov-Arnold Network (Wav-KAN) for federated learning, leveraging the strengths of wavelet transformations to enhance neural network performance. Through extensive experimentation on various datasets, including MNIST, we demonstrated that Wav-KAN outperforms traditional methods in both training and test accuracy. By incorporating wavelets such as Mexican hat, Morlet, Derivative of Gaussian (DOG), and Shannon wavelets, Wav-KAN effectively captures multi-resolution features, leading to improved computational efficiency and robustness.

Our federated learning algorithm integrates wavelet-based activation functions parameterized by weight, scale, and translation, facilitating effective local and global model updates. The results highlight the significant impact of wavelet selection on the model's performance, emphasizing the superiority of certain wavelets in handling non-stationary signals and localized features.

Overall, Wav-KAN presents a scalable and efficient approach for federated learning, addressing the limitations of traditional methods and offering a robust framework for neural network design. We are working on unbalanced and non-i.i.d data distributions across various datasets, further validating the model's effectiveness and convergence properties.

\ifCLASSOPTIONcaptionsoff
  \newpage
\fi

\bibliographystyle{IEEEtran}
\bibliography{references}

\begin{thebibliography}{1}
\providecommand{\url}[1]{#1}
\csname url@samestyle\endcsname
\providecommand{\newblock}{\relax}
\providecommand{\bibinfo}[2]{#2}
\providecommand{\BIBentrySTDinterwordspacing}{\spaceskip=0pt\relax}
\providecommand{\BIBentryALTinterwordstretchfactor}{4}
\providecommand{\BIBentryALTinterwordspacing}{\spaceskip=\fontdimen2\font plus
\BIBentryALTinterwordstretchfactor\fontdimen3\font minus
  \fontdimen4\font\relax}
\providecommand{\BIBforeignlanguage}[2]{{%
\expandafter\ifx\csname l@#1\endcsname\relax
\typeout{** WARNING: IEEEtran.bst: No hyphenation pattern has been}%
\typeout{** loaded for the language `#1'. Using the pattern for}%
\typeout{** the default language instead.}%
\else
\language=\csname l@#1\endcsname
\fi
#2}}
\providecommand{\BIBdecl}{\relax}
\BIBdecl

\bibitem{wav-kan}
Z.~Bozorgasl and H.~Chen, ``Wav-kan: Wavelet kolmogorov-arnold networks,''
  \emph{arXiv preprint arXiv:2405.12832}, 2024.

\bibitem{witkin}
A.~P. Witkin, ``Scale-space filtering,'' in \emph{Readings in computer
  vision}.\hskip 1em plus 0.5em minus 0.4em\relax Elsevier, 1987, pp. 329--332.

\bibitem{mallat}
S.~Mallat, \emph{A wavelet tour of signal processing}.\hskip 1em plus 0.5em
  minus 0.4em\relax Elsevier, 1999.

\end{thebibliography}

\end{document}